\documentclass[twocolumn]{aastex63}
\usepackage{booktabs}
\usepackage{graphicx}
\usepackage{amssymb}
\usepackage{lipsum} 

\makeatletter
\newcommand*{\rom}[1]{\expandafter\@slowromancap\romannumeral #1@}
\makeatother
\received{17 November, 2020}
\revised{27 January, 2021}
\accepted{7 February, 2021}
\submitjournal{ApJ}

\shorttitle{Multi-wavelength pulsations via time-dependent particle acceleration }
\shortauthors{Clarke et al.}
\graphicspath{{./}{figures/}}
\begin{document}

\title{Quasi-Periodic Particle Acceleration in a Solar Flare}

\correspondingauthor{Brendan P. Clarke}
\email{clarkeb3@tcd.ie}

\author[0000-0003-1065-8167]{Brendan P. Clarke}
\affiliation{School of Cosmic Physics, Dublin Institute for Advanced Studies, Dublin, D02 XF85, Ireland}
\affiliation{School of Physics, Trinity College Dublin, Dublin 2, Ireland}

\author[0000-0002-6835-2390]{Laura A. Hayes}
\affiliation{Solar Physics Laboratory, Heliophysics Science Division, NASA Goddard Space Flight Center, Greenbelt, MD 20771, USA}

\author[0000-0001-9745-0400]{Peter T. Gallagher}
\affiliation{School of Cosmic Physics, Dublin Institute for Advanced Studies, Dublin, D02 XF85, Ireland}

\author[0000-0002-4715-1805
]{Shane A. Maloney}
\affiliation{School of Cosmic Physics, Dublin Institute for Advanced Studies, Dublin, D02 XF85, Ireland}
\affiliation{School of Physics, Trinity College Dublin, Dublin 2, Ireland}

\author[0000-0002-6106-5292]{Eoin P. Carley}
\affiliation{School of Cosmic Physics, Dublin Institute for Advanced Studies, Dublin, D02 XF85, Ireland}
\affiliation{School of Physics, Trinity College Dublin, Dublin 2, Ireland}



\begin{abstract}
A common feature of electromagnetic emission from solar flares is the presence of intensity pulsations that vary as a function of time. Known as quasi-periodic pulsations (QPPs), these variations in flux appear to include periodic components and characteristic time-scales. Here, we analyse a GOES M3.7 class flare exhibiting pronounced QPPs across a broad band of wavelengths using imaging and time-series analysis. We identify QPPs in the timeseries of X-ray, low frequency radio and EUV wavelengths using wavelet analysis, and localise the region of the flare site from which the QPPs originate via X-ray and EUV imaging. It was found that the pulsations within the 171~\AA, 1600~\AA, soft X-ray (SXR), and hard X-ray (HXR) light curves yielded similar periods of $122^{+26}_{-22}$ s, $131^{+36}_{-27}$ s, $123^{+11}_{-26}$ s, and $137^{+49}_{-56}$ s, respectively, indicating a common progenitor. The low frequency radio emission at 2.5 MHz contained a longer period of ${\sim}231$ s. Imaging analysis indicates that the location of the X-ray and EUV pulsations originates from a HXR footpoint linked to a system of nearby open magnetic field lines. Our results suggest that intermittent particle acceleration, likely due to `bursty' magnetic reconnection, is responsible for the QPPs. The precipitating electrons accelerated towards the chromosphere produce the X-ray and EUV pulsations, while the escaping electrons result in low frequency radio pulses in the form of type \rom{3} radio bursts. The modulation of the reconnection process, resulting in episodic particle acceleration, explains the presence of these QPPs across the entire spatial range of flaring emission.
\end{abstract}


\section{Introduction} \label{sec:intro}

Quasi-periodic pulsations (QPPs) are an important feature observed in solar and stellar flare emission \citep{Nakariakov2009, doors, newreview}. This puzzling phenomenon lacks a concrete definition, however, they are typically described by variations in the flux from a flare as a function of time that appear to include periodic components and time-scales that typically range from 1 s up to 1 min, and in extreme cases from $\le$1 s up to several minutes \citep{karleeky, Tan_2010, Li_2015, hayes2019}. QPPs are typically observed during the impulsive phase of solar flares, however, in recent years it has become clear that they can persist through to the decay phase, after the impulsive energy release \citep{Hayes2016a, Dennis2017, hayes2019}.
\begin{figure}[t!]
\centering
\includegraphics[width=8cm]{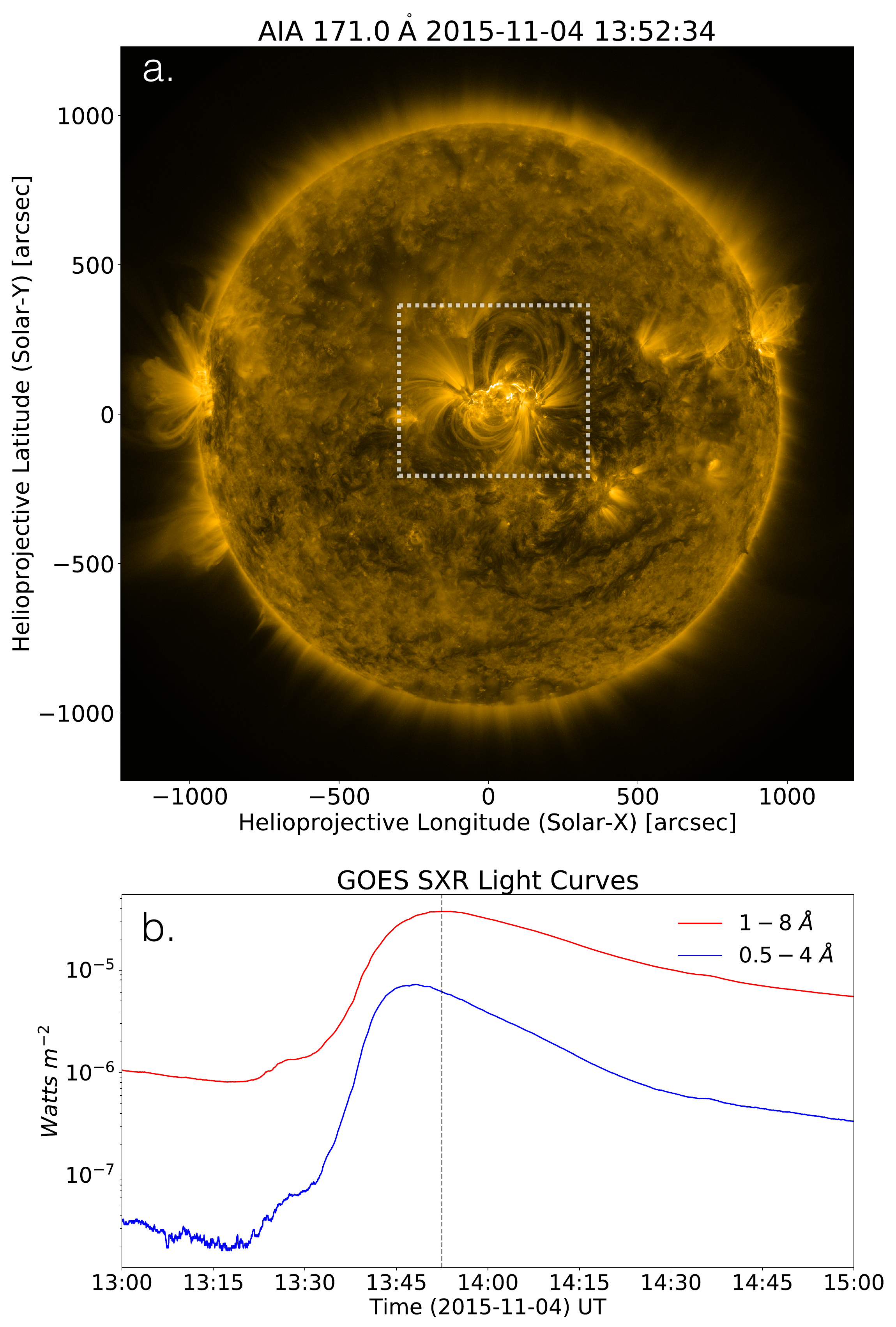}
\caption{(a): SDO/AIA 171~\AA\ passband image of the sun on 2015 Nov 4. The active region associated with the event is visible in the dashed box at the disk center. (b): The GOES SXR light curves showing the occurrence of the M3.7 class flare. The dashed grey line shows the time at which the image in panel a was taken.}
\label{fig1}
\end{figure}

QPPs have been reported in a broad range of wavelengths from decametric radio \citep{Li_2015, carley2019}, through to extreme ultra-violet (EUV) and X-rays, \citep{dom, dolla}, and even $\gamma$ rays \citep{Nakariakov_2009_a, li2020}. Statistical studies suggest that QPPs are a common feature, especially in larger flaring events \citep{Simoes2015, inglis2016, hayes2020}. Within the decimetric waveband, QPPs can manifest as a sequence of type \rom{3} radio bursts emanating from the corona as a consequence of accelerated beams of electrons escaping along open magnetic field lines away from the flare site \citep{ashwandenb, ning, Kupriyanova2016}. In contrast, QPPs in the EUV are typically observed to originate from the hot plasma in the coronal loops of a flaring region \citep{doors}. In addition to studies of QPPs analysed within specific spectral domains, some research has been done focusing on events containing QPPs across a wide band of wavelengths. For example, \cite{Aschwanden1993} investigated the timing of HXR pulsations with respect to pulsations seen in radio wavelengths (100-300 MHz) and found evidence for a strong causal connection. Additionally, \cite{Tajima1987} found that current loop coalescence can lead to quasi-periodic amplitude oscillations in the microwave, X-ray, and $\gamma$-ray wavebands. More recently, \cite{kumar2016} presented a multi-wavelength analysis of QPPs found to be occurring in HXR, radio (25–180, 245, 610 MHz), and EUV wavelengths. 

Several models have been proposed as explanations for the presence of QPPs in solar and stellar flares \citep{mclaughlin} which are typically categorised as oscillatory or self-oscillatory processes. In the regime of oscillatory processes, QPPs are interpreted as a signature of magnetohydrodynamic (MHD) oscillations inducing periodic motions about an equilibrium in the flaring region. This explanation has been promising for some events, as some observed periodicities of QPPs are in good agreement with that of the timescales of MHD waves in the corona \citep{Nakariakov2009}. There is widespread observational evidence for MHD waves existing in the corona and it is possible that kink, toroidal, longitudinal, or sausage modes could cause some of the thermal and non-thermal intensity variations that we observe. 

For example, kink mode oscillations have been reported that have an overlapping timescale ($\sim$1.5-10 minutes) with observed QPP periodicities \citep{Anfinogentov_2015}. Such waves could periodically modulate emission or influence particle dynamics \citep{Nakariakov2009}. It is also possible that the presence of these waves could trigger magnetic reconnection resulting in a periodicity related both to the type of wave mode involved and the properties of the coronal loops \citep{zim, carley2019}.

In the other category of self-oscillatory processes, QPPs are interpreted as a manifestation of time dependent, intermittent magnetic reconnection. Specific mechanisms theorised to produce intermittently accelerated electron beams resulting in QPPs that are relevant to this work include plasmoid magnetic island reconnection and oscillatory reconnection (reconnection reversal). For example, \cite{Kliem2000} demonstrated in a 2D numerical model of a long current sheet that the formation of plasmoids and their coalescence could result in quasi-periodic particle acceleration, which they used to explain the presence of decimetric radio bursts with periods of 0.5-10s. In a more recent study, \cite{Guidoni2016} built upon the work of \cite{Drake2006} to show that the generation, contraction, and interaction of magnetic islands in a multi-layered current sheet can efficiently accelerate charged particles. It was demonstrated that this mechanism should occur in a manner that is sporadic and intermittent, and hence would result in pulsating flare emission. The period in this case would be related to the rate of island formation and their interaction with the flaring arcade. 

In addition to plasmoid-dominated reconnection models, numerical simulations have demonstrated that the reconnection process itself can be oscillatory. In this scenario of oscillatory reconnection, competition between the thermal-pressure gradients and the Lorentz force provide a restoring force as each aspect overshoots the other in search for equilibrium \citep[e.g.][]{mclaughlin09, murray09}. Simulations have shown that this mechanism can produce oscillatory reconnection which results in intermittent particle acceleration in a self-consistent manner resulting in decaying QPPs with periodicities of $\sim$105–212.5 s \citep{McLaughlin2012Oscillatory}. This work has also been extended to a 3D null-point, and it has been shown that reconnection can naturally proceed in a time-dependent oscillatory behaviour \citep{thurgood_2017}.

Despite a plethora of studies and observations of QPPs in various contexts, the underlying mechanism for their generation remains a topic of debate. The challenge in identifying an underlying mechanism is that the QPPs are linked to many aspects of flaring emission, and encompass electromagnetic emission that originates from the very base HXR footpoints of a flare up to altitudes of several solar radii in the corona. Hence detailed studies that identify QPPs from multiple wavelengths are required to connect different aspects of the flaring process in order to constrain the mechanism producing them. In this paper we present a study that demonstrates prominent QPPs occurring over an unusually wide range of frequencies. We observe non-thermal emission in the form of HXRs produced via bremsstrahlung in the chromosphere, co-temporal thermal emission via SXR and EUV emission in the flare loops/transition region, as well as a sequence of low frequency type \rom{3} bursts emanating from the high corona. We analyse the multiple types of emission mechanisms at play including thermal emission, plasma emission, and non-thermal bremsstrahlung. These mechanisms are all associated with the same intermittently accelerated electron beams which result in the pulsations that we can localise to a specific region of the flare site. This region is associated with a system of open and closed field lines. Identifying the source region responsible for the intermittent acceleration of the electron beams is novel from the perspective of analysing QPPs. This work provides new observational evidence that QPPs can originate from an identifiable specific regions of flares and manifest across the entire electromagnetic spectrum via multiple emission mechanisms. The QPPs observed in the radio regime are notably lower in frequency compared to what is typically observed. This indicates the considerable distances over which QPPs can manifest - from the solar chromosphere through to interplanetary space (${\sim}16$ $R_\odot$)

 \begin{figure}[t!]
\centering
\includegraphics[width=8.5cm]{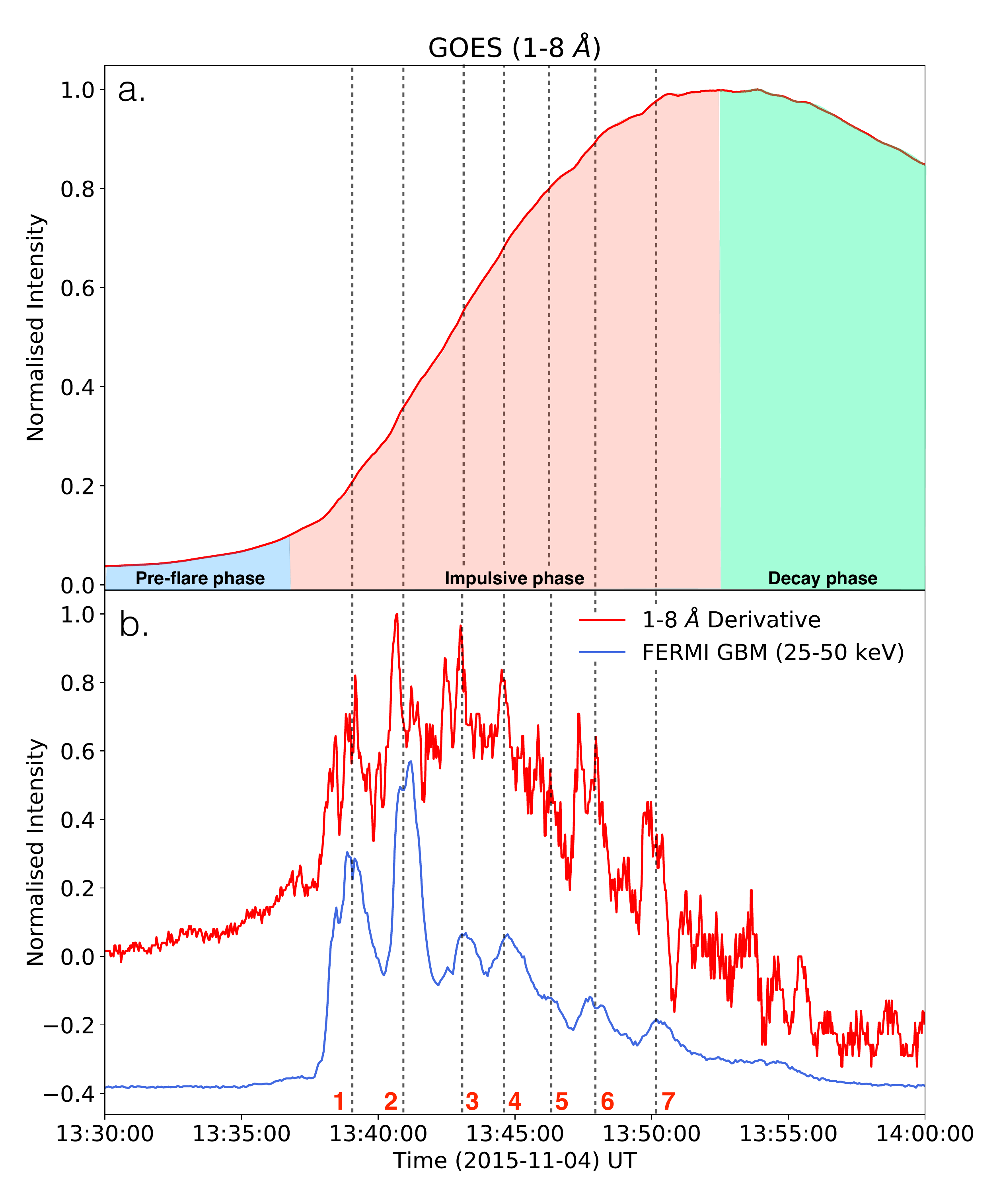}
\caption{(a): The normalised SXR light curve from GOES (1-8~\AA) at the time of the flare. (b): The time derivative of the SXR emission and the HXR light curve from FERMI GBM (25-50 keV). The QPPs present are labelled one through seven. Here, we see a clear illustration of the Neupert effect and indicate the seven primary QPPs analysed in this work.}
\end{figure}
The flare presented in this paper is a Geostationary Operational Environmental Satellite (GOES) M3.7 class flare that occurred on 04-Nov-2015. An overview of the active region located at the center of the solar disk as observed in 171~\AA\ is shown in Figure~\ref{fig1}a and the light curves of the flare from the GOES X-Ray Sensor (XRS) in two channels (1-8~\AA\ in red and 0.5 - 4~\AA\ in blue) is shown in Figure~\ref{fig1}b. In Section 2, the instruments used for this study: The Atmospheric Imaging Assembly (AIA) on board the Solar Dynamics Observatory (SDO) \citep{lemen2012}, the Reuven Ramaty High Energy Solar Spectroscopic Imager \citep{lin2002}, the WAVES instrument on board the WIND satellite \citep{wind}, The Gamma-ray Space Telescope (GBM) onboard Fermi \citep{meegan2009}, and the GOES XRS are briefly introduced. Details of the analysed event and data analysis techniques are also included in this section. In Section 3, we present our observations alongside our analysis of the QPPs. In Section 4, we present a discussion and interpretation of the work before concluding our findings in Section 5.
\section{Instrumentation, observations, and data analysis}
\subsection{Instrumentation and observations}
The GOES XRS measures the solar SXR fluxes integrated over the entire solar disk. It has a cadence of 2 s with two channels of 0.5-4~\AA\ and 1-8~\AA. In this work, we primarily focus on the 1-8~\AA\ channel as it exhibited the most pronounced QPPs. Figure 2a shows this light curve. The pre-flare, impulsive, and decay phases are also indicated. The event began at 13.31 UT and peaked at 13.52 UT. Figure 2b shows the time derivative of this light curve with the HXR light curve observed by FERMI GBM (25-50 keV) overplotted. The Neupert effect, which refers to the observed phenomenon that non-thermal HXR emission coincides temporally with the rate of rise of the thermal SXR emission (i.e. the derivative), is observed here as it is clear that the pulsations in the SXR derivative are coincidental with those observed in the HXR emission \citep{Neupert_1968}. This relates the HXR flux from the flare `footpoints' to the thermal SXRs observed from the heated plasma. We identify seven 
distinct pulsations throughout the event as shown in Figure 2b. One can see that these pulsations all occur during the impulsive phase of the flare. This indicates that the mechanism producing these QPPs must be able to modulate the acceleration of electrons. However, it is clear that some pulsations do persist into the decay phase within the SXR emission. 
\begin{figure}[b!]
\centering
\includegraphics[width=8cm]{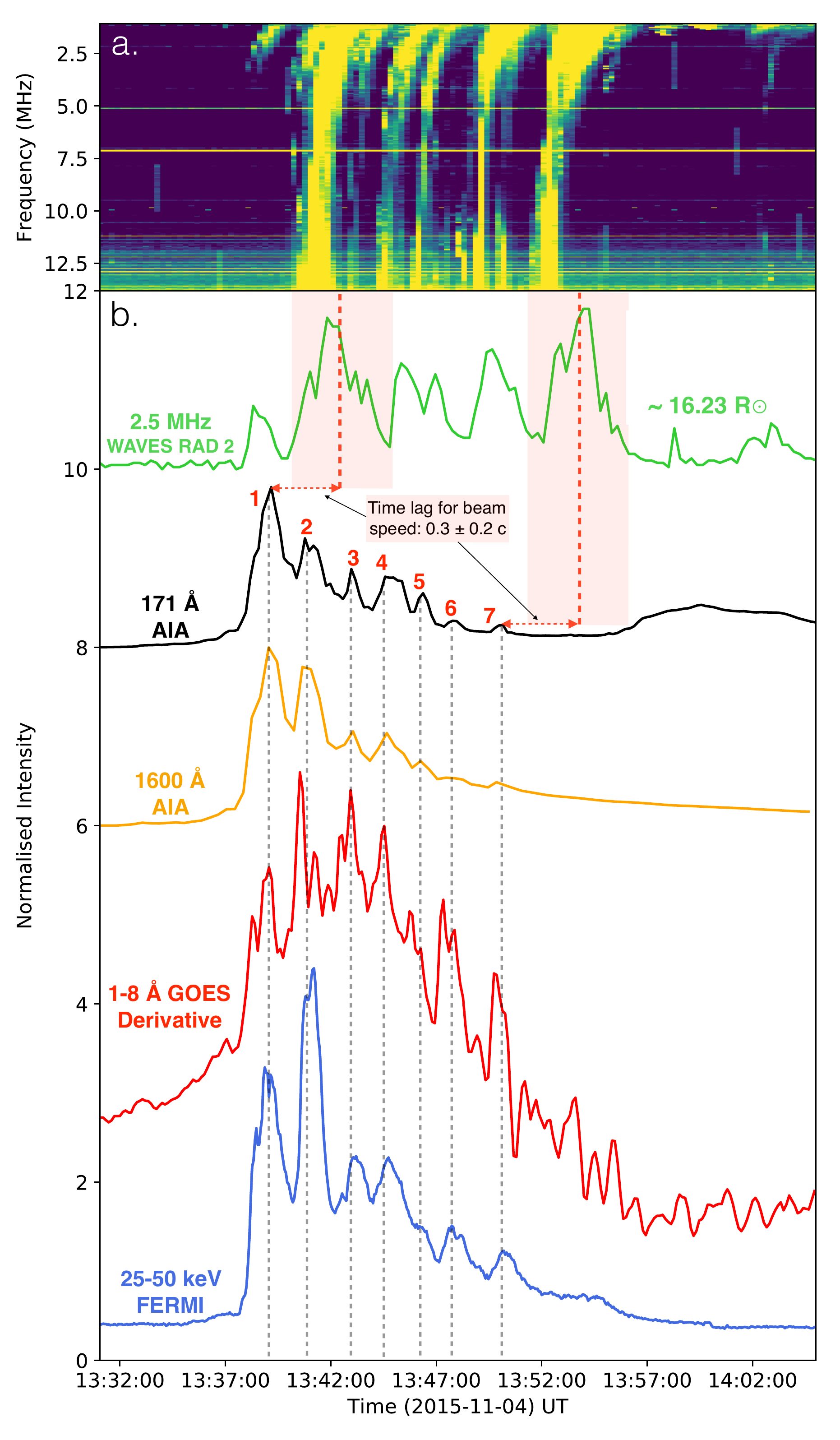}
\caption{(a): Dynamic spectrum from WIND/WAVES RAD2 showing a series of pulsed type \rom{3} radio bursts. (b): Multi-wavelength light curves observed from a number of instruments. From top to bottom the lightcurves go from longer to shorter wavelenghts. This is a proxy for altitude of the flaring emission source with the radio data representing emission originating from high in the corona down to the HXR emission originating from the footpoints. The black, orange, red, and blue light curves have QPPs well correlated in time. The green light curve shows the radio emission at 2.5 MHz. This frequency was chosen as it captured each pulse most effectively. The radio QPPs have a longer periodicity which we elaborate on in the discussion section. Lines drawn from pulses 1 and 7 show the time lag needed for the source electron beams of typical speeds to reach $16$ $R_\odot$ (approximate height of the radio source) from the flare site.}
\end{figure}

The most pronounced pulsations we observed with AIA were from the 171~\AA\ and 1600~\AA\ passbands. Images from these passbands were used to analyse the periodicity and spatial distribution of the QPPs in EUV. The cadences of these images were 12~s and 24~s, respectively. The 171~\AA\ passband is dominated by the Fe \rom{11} line and most represents emission from the corona and upper transition region while the 1600~\AA\ passband is dominated by C \rom{4} and images primarily the upper photosphere/transition region \citep{aia}. These images enable us to estimate the layer of the atmosphere from which the QPPs we observe originate and how they relate to one another. 

RHESSI observed the event up until 13:43 UT before entering spacecraft night. This allowed us to image the location of the HXRs produced during the flare for our analysis. Although RHESSI was unable to observe all the HXR emission throughout the flare, Fermi GBM captured this information which we incorporate into our analysis. The WIND/WAVES RAD2 instrument was used to gather radio data. Dynamic spectra from 0.02-13.85 MHz (cadence: 16.188 s) were analysed to investigate the low-frequency aspect of the QPPs in the event. The emission at these wavelengths manifests in the form of type \rom{3} radio bursts which are a result of plasma emission \citep{reid2014}. Within this mechanism, the frequency, $f_{p}$, of the radiation is proportional to the local electron density $n_{e}$ via $f_{p}\sim8980\sqrt{n_e}$. The electron density in the corona decreases with height. Hence, by using an electron density model, the height of a radio source produced through plasma emission may be estimated. At the frequency of 2.5 MHz, most of these bursts are captured and so we use this frequency to generate a light curve of the radio data. Using the electron density model from \cite{newkirk1967}, this corresponds to a height of ${\sim}16$ $R_\odot$. 

Together, we can use these data to determine information about the periodicity and location of the QPPs observed from the HXR footpoints through to the upper transition region and corona. The altitude at which each different waveband emits differs significantly. For example, the HXRs are produced through bremsstrahlung by non-thermal electrons colliding with the chromosphere while the type \rom{3} radio bursts are produced via plasma emission. Figure 3 is summary of the QPPs analysed in this work. Figure 3a shows the dynamic spectrum of the radio emission containing a sequence of type \rom{3} radio bursts and Figure 3b shows the EUV, SXR and HXR light curves in which we identify 7 distinct QPPs. The EUV light curves were extracted from the QPP source region we identified which is explained in Section 3.2.
\subsection{Data analysis and imaging}
Wavelet analysis using a Morlet wavelet was performed on the multiwavelength light curves to determine their periodicities using the software\footnote{http://atoc.colorado.edu/research/wavelets/} developed by \cite{terrence1998}. This technique is a powerful tool for searching within time-series for periodic signatures as, unlike Fourier analysis, it provides a 2D spectrum of both frequency and time allowing one to assess if a quasi-periodic signature varies in time \citep{moortel}.

\begin{figure}[t!]
\centering
\includegraphics[width=8.5cm]{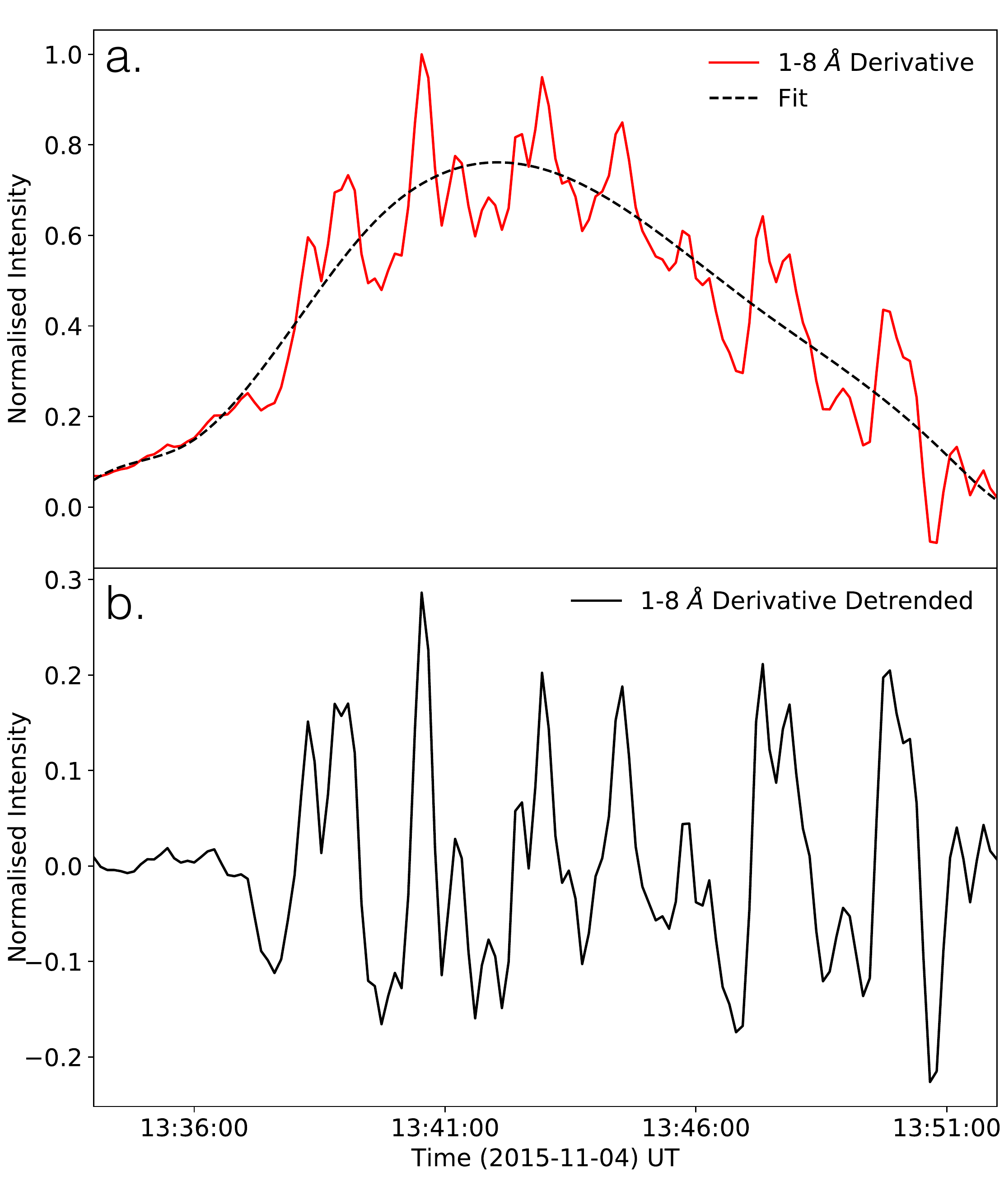}
\caption{Example of the background subtraction technique used to isolate the QPPs. (a): The time derivative of the SXR emission is shown in red with a spline fit to the overall large-scale slowly varying emission overplotted in black . The fit excludes the shorter time-scale variation of the QPPs. (b): The subtraction of the fit from the SXR emission resulting in a time-series containing the QPPs without the slowly varying background emission.}
\end{figure}

In order to more accurately determine the period of the QPPs via the wavelet analysis, the slowly varying background trend of the flare emission was removed from each time-series. This process is shown in Figure 4 for the case of the SXR emission from GOES. To do this, a spline fit was constructed using a 3rd order polynomial with 28 samples between nodes. The fit was then subtracted from the original data resulting in a time-series containing only the modulation of the emission resulting from the QPPs. This process was repeated for each time-series we analysed. No subtraction was required for the radio light-curve at 2.5 MHz as there was little background in this data. Care was taken to ensure each fit accurately represented the slowly varying background emission in order to avoid introducing any artefacts during the subtractions.

In addition to carrying out the wavelet analysis on the multi-wavelength detrended time-series, we also performed the same analysis on the relevant data without detrending in order to cross-check our results. This analysis is provided in the appendix of this paper. We also manually determined the period of the QPPs. This was achieved by visually identifying the time of each pulse and plotting these times against pulse number (1-7). The period can then be simply estimated by fitting a straight line to this data and finding the slope.

The PIXON algorithm was used to image the RHESSI HXR sources. It seeks a superposition of circular sources of different sizes and parabolic profiles that most replicate the modulations measured by the detectors, while maintaining the fewest degrees of freedom possible. PIXON is thought to provide accurate image photometry in comparison to the other faster algorithms such as CLEAN \citep{rhessi}. Images taken by SDO/AIA were used to analyse the most prominent pulsations in the EUV regime which were found in the 171~\AA\ and 1600~\AA\ passbands. Time-series were constructed from these images by integrating the emission over various regions of the flare in order to localise the area producing the pulsations. This is discussed further in Section 3.2. Additional data analysis was carried out to estimate the height of the source producing the radio emission via the \cite{newkirk1967} electron density model. This height was determined to be ${\sim}16$ $R_\odot$. Figure 3b shows the time lag required to reach this height from the flaring region with beam speeds of 0.1-0.5 c. Type \rom{3} radio bursts typically have source electron beam velocities of $\sim$0.3 c but have been found to vary from 0.1-0.5 c in some cases \citep{reid2014}.
\begin{figure*}[t!]
\centering
\includegraphics[width=15cm]{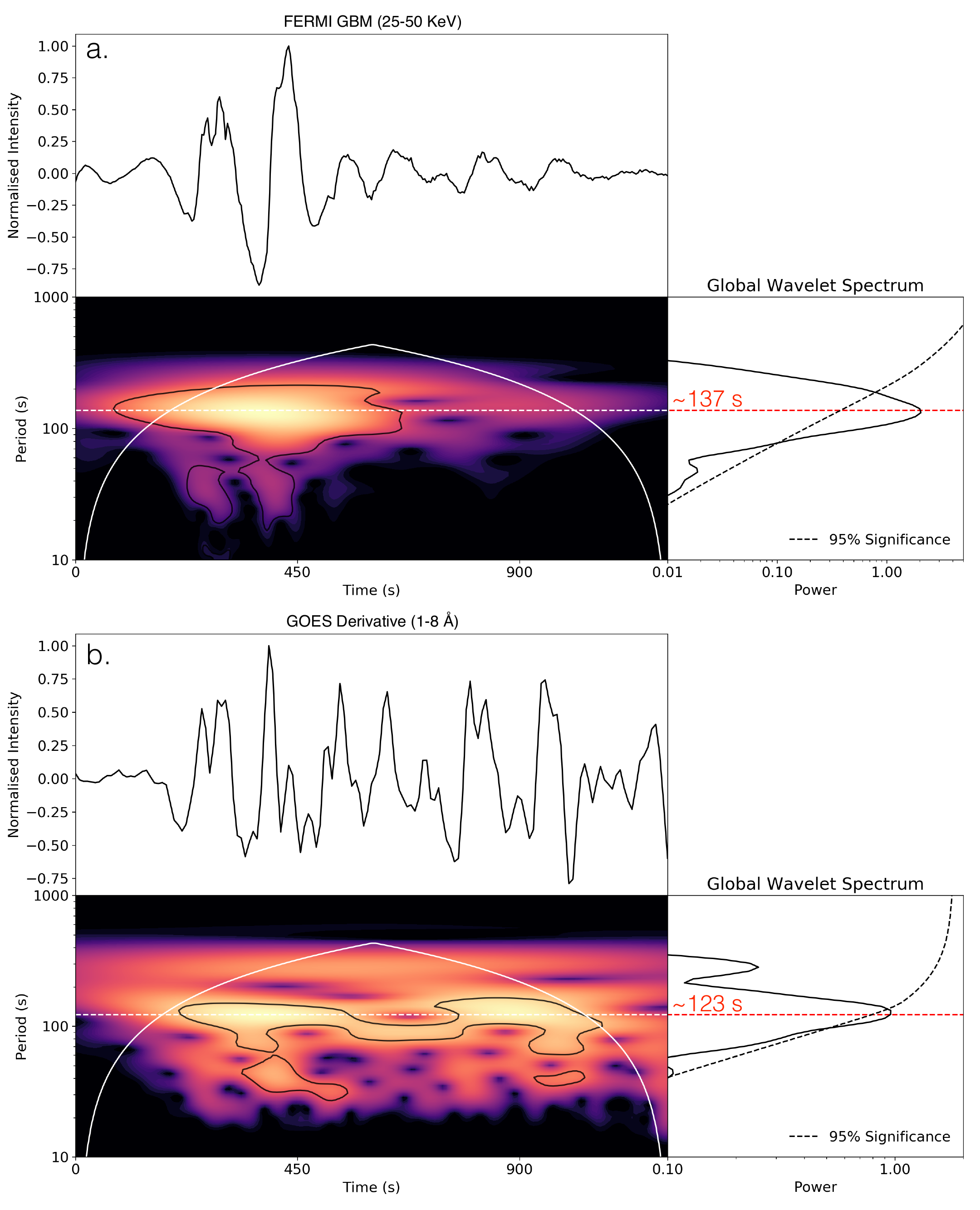}
\caption{Wavelet analysis of the detrended (a): HXR and (b): SXR derivative emission from the flare. The periods were found to be $137^{+49}_{-56}$ s and $123^{+11}_{-26}$ s, respectively. The error is taken as the range over which each global power spectrum is above 95\% significance.}
\end{figure*}
\begin{figure*}[t!]
\centering
\includegraphics[width=15cm]{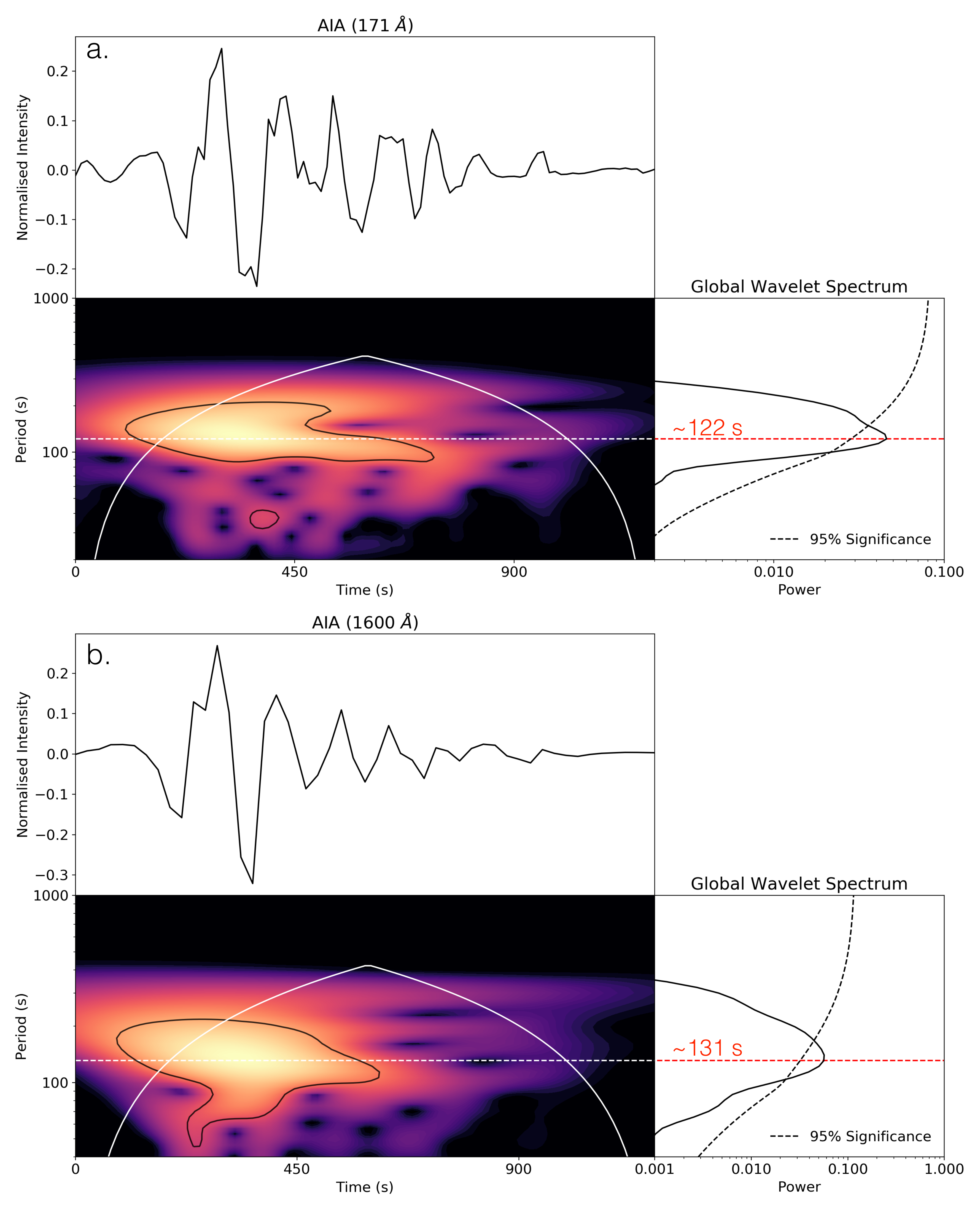}
\caption{(a): Wavelet analysis of the detrended emission at 171~\AA. (b): Wavelet analysis of the emission at 1600~\AA. The periods were found to be $122^{+26}_{-22}$ s and $131^{+36}_{-27}$ s respectively.}
\end{figure*}
\section{Results}
Across the electromagnetic spectrum, the impulsive nature of the event begins at $\sim$13.37 UT and continues until $\sim$13.57 UT. Figure 3b shows the normalised lightcurves under investigation that spans from the HXR to the low frequency radio emission, all of which contain QPPs. The lightcurves were normalised by dividing each time series by its peak value. They were then scaled in order to plot them in a vertical sequence for ease of comparison, going from higher frequency to lower frequency (bottom to top). The HXR light curve (25-50 keV) from Fermi/GBM in blue is at the bottom of Figure 3b, with the most prominent pulsations labelled one through seven. 
\begin{figure*}[t!]§
\centering
\includegraphics[width=15cm]{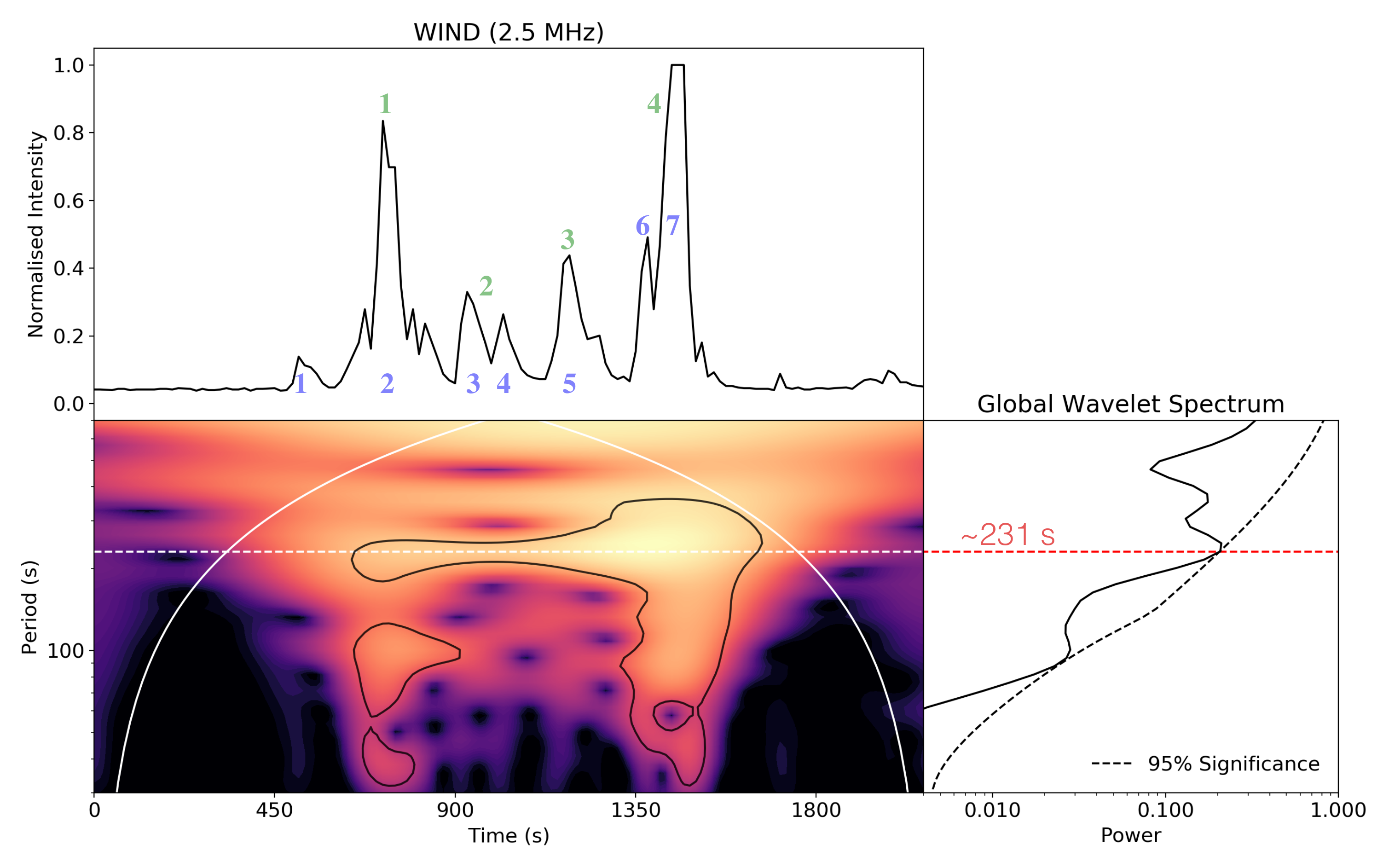}
\caption{Wavelet analysis of the radio emission at 2.5 MHz. The period was found to be ${\sim}231$ s.}
\end{figure*}
Above in red, is the derivative of the GOES light curve (1-8~\AA), followed by the 1600~\AA~curve in orange, and the 171~\AA~curve in black. The clear co-temporal presence of the pulsations in each of these light curves obtained by simultaneous observations from different instruments makes clear that these QPPs are of solar origin and are not due to some instrumental effect.
\begin{figure}[t]
\centering
\includegraphics[width=8.7cm]{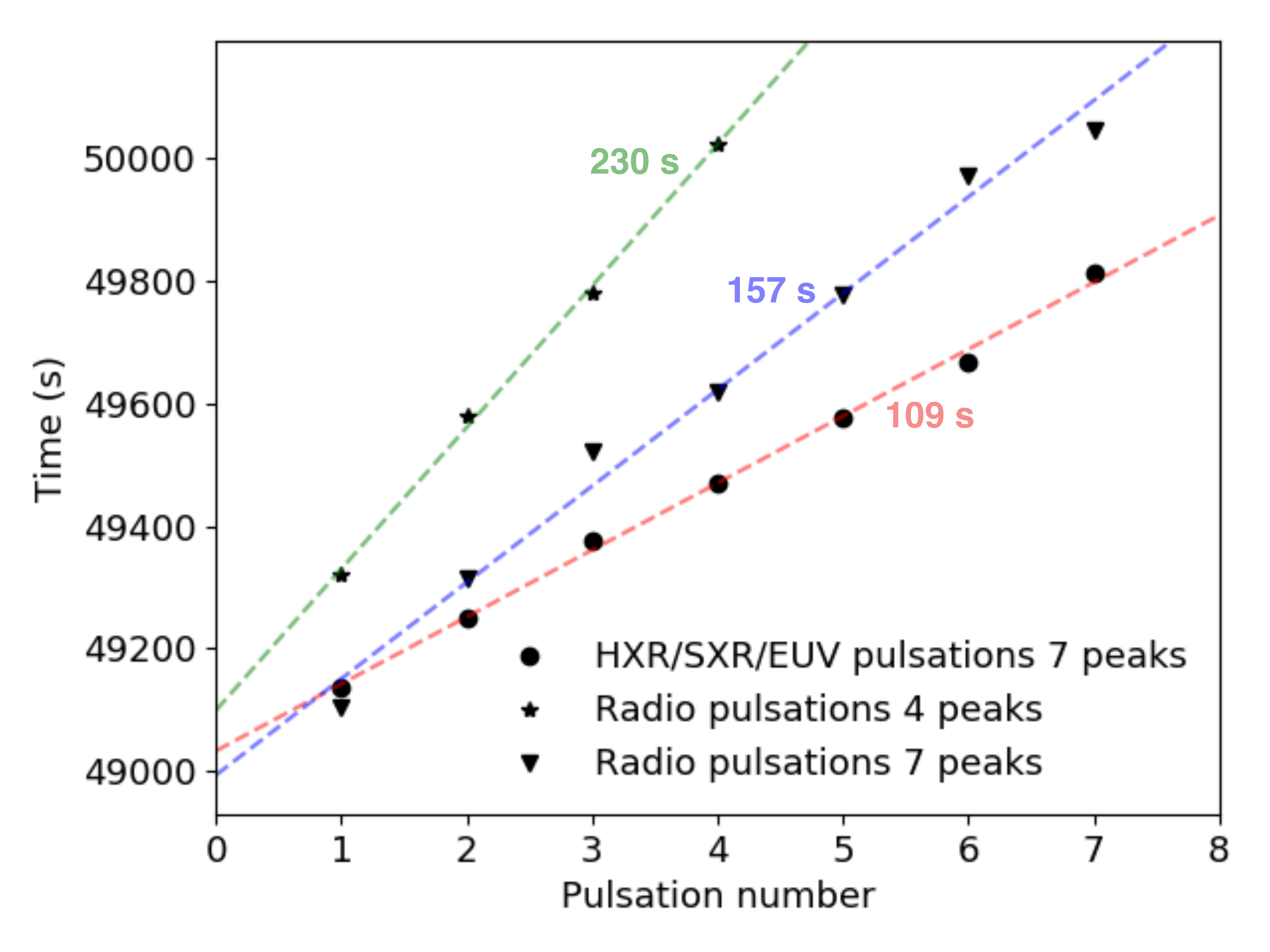}
\caption{Pulse number versus time of each pulse. The 7 HXR/SXR/EUV pulses shown in Figure 3 are plotted using the circle symbols. The slope of the straight line fitted to the data provides an estimate of the period: $\sim 109$ s. For the radio emission at 2.5 MHz, this analysis was done for the four main peaks in the time-series, as well as for seven peaks in the case where lower amplitude peaks are included. This resulted in periods of $\sim 230$ s and $\sim 157$ s, respectively.}
\end{figure}
Figure 3a shows the dynamic spectrum from WIND/WAVES RAD2 (1.075–13.825 MHz) that contains a series of pulsed type \rom{3} radio bursts during the same time frame in which the QPPs were observed. The green light curve in Figure 3b shows a slice from this dynamic spectrum at 2.5 MHz. This frequency was chosen as it contained the majority of the bursts. Using the \cite{newkirk1967} electron density model, we can estimate the height at which this radio emission is emitted: ${\sim}16$ $R_\odot$. Lines drawn from the peaks of pulses 1 and 7 from the 171~\AA~curve are shown that indicate the time delay required for electron beams of velocities between 0.1 and 0.5 c to reach this height. For both pulse 1 and 7, there appears to be radio pulsations that occur at the expected times. This analysis suggests that the electron beam speeds are close to the typical value of 0.3 c for type \rom{3} radio bursts. It is clear that the radio QPPs at 2.5 MHz are less correlated with the higher frequency radiation. There are a number of reasons for which one would not expect a one-to-one relation between radio pulsations produced via plasma emission in interplanetary space and the higher frequency emission produced via bremsstrahlung/heating close to the flare site despite originating from the same populations of accelerated electrons. These differences are elaborated upon in the discussion section. 
\subsection{Periodicities}
For each lightcurve, wavelet analysis was conducted over the same time period: 13:34-13:54 UT. The error for each result was taken as the range over which the global power spectrum was above the 95\% significance curve. The analysis was carried out on the detrended light curves. However, the appendix includes the same analysis for the data without detrending. The results agreed in both cases.
\begin{figure*}[t!]
\centering
\includegraphics[width=18cm]{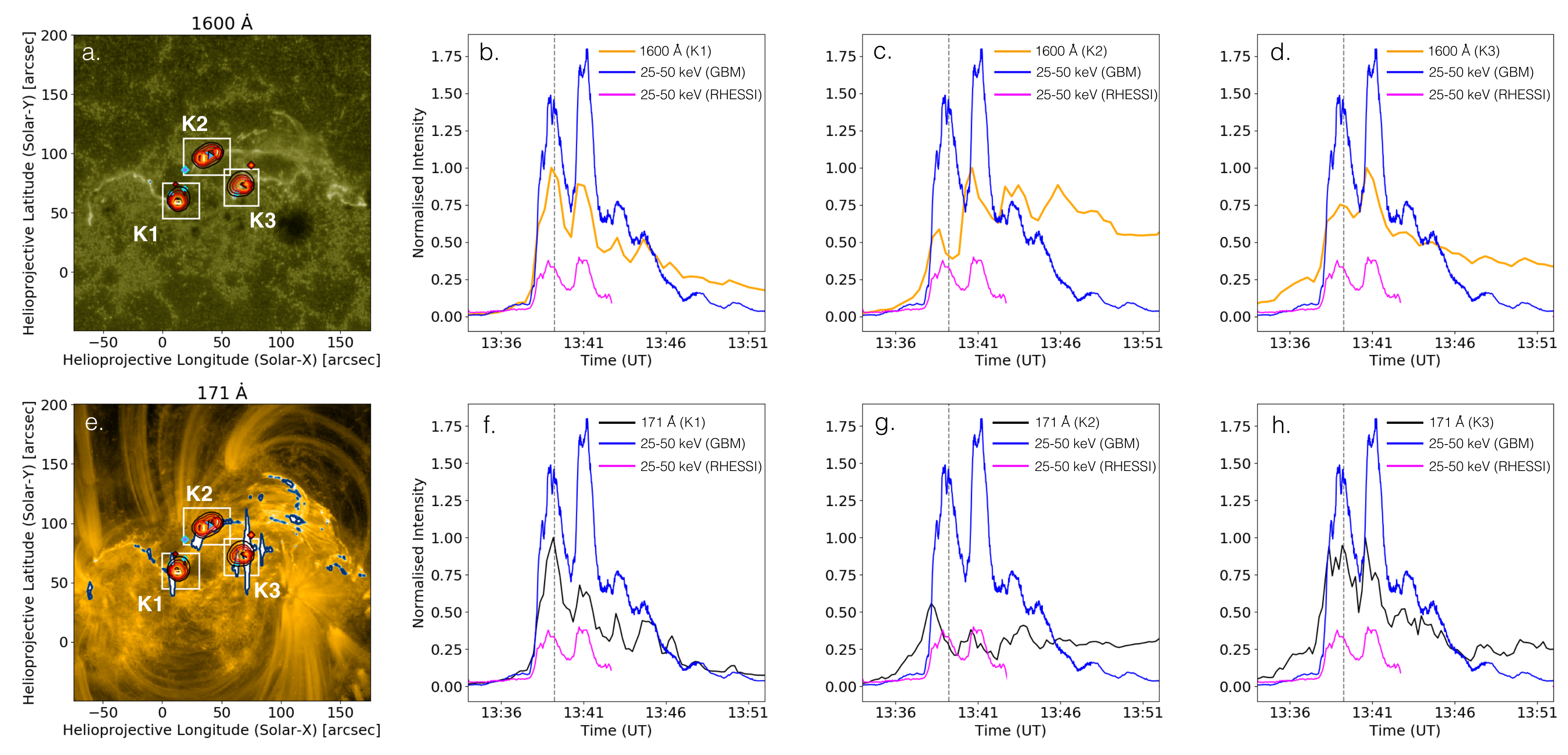}
\caption{Spatial analysis of QPPs. (a): Image of the flaring region in 1600~\AA~taken at the time of the first QPP. The RHESSI image of the three HXR footpoints is overlaid which are labelled within three kernels as K1, K2, and K3. K1 and K2 lie along one flare ribbon while K3 is located on the other. This image was constructed using the PIXON algorithm over the available time period during which RHESSI captured the emission. This time period is shown in magenta with the time-series data in (b). (b-d): Light curves of the HXRs observed by GBM and RHESSI (25-50 keV) and the light curves extracted from the 1600~\AA\ images taken by AIA at each HXR source location. The light curves were constructed by integrating over the pixels contained in the boxes surrounding the HXR sources in (a). The same analysis is done for the 171~\AA\ images as shown in (e-h). It was found that the emission in EUV from within K1 produced light curves containing QPPs most correlated with those seen in the HXR emission (b+f). This localises the source of the QPPs to this region of the flare site which is close to a nearby system of open and closed magnetic field lines. The animation related to this Figure, provided in the online version of this article, shows the evolution of the flare at each time step from 13:34 - 13:53 UT.}
\end{figure*}

Figure 5a shows the wavelet analysis that was carried out on the GBM 25-50 keV lightcurve. In this plot, the HXR time-series, the wavelet power spectrum, illustrating power at particular periodicities as a function of time, and the global power spectrum are shown. A period of $137^{+49}_{-56}$ s was found in this channel. Figures 5b, 6, and 7 show this same analysis for the SXR, EUV, and radio wavelengths, respectively. The SXR emission contained a period of $123^{+11}_{-26}$ s, while the pulsations within the 171~\AA~and 1600~\AA~light curves yielded periodicities of $122^{+26}_{-22}$ s and $131^{+36}_{-27}$ s, respectively. The 2.5 MHz light curve was found to have significant period at a time-scale of ${\sim}231$ s. The time-scales of the QPPs in in the 171~\AA, 1600~\AA, SXR (1-8~\AA), and HXR (25-50 keV) are therefore all in good agreement within error. These results are summarised in Table 1.

In addition to calculating the periods of the light curves via wavelet analysis, we also estimated them manually by visually identifying peaks. Figure 8 shows the time of the HXR, EUV, and SXR pulsations versus pulse number (see pulses 1-7 in Figure 3). The slope of this line provides an estimate of the period. The result was found to be ${\sim}109$ s. 
\begin{table}[b]
\scriptsize
\centering
\caption{Comparison of the periods found in the lightcurves for each analysed wavelength via wavelet analysis. The cadence of the data in each case, $\Delta t$, is also shown.}
\label{tab:my-table}
\begin{tabular}{ c c c c c c }
\toprule
$\lambda$ &2.5 MHz&171 \AA&1600 \AA&1-8 \AA&25-50 keV\\ \midrule
Period &$\sim$231 s&$122^{+26}_{-22}$ s&$131^{+36}_{-27}$ s&$123^{+11}_{-26}$ s&$137^{+49}_{-56}$\\
$\Delta t$ &16.2 s&12 s&24 s&2 s&1.6\\ \bottomrule
\end{tabular}%
\end{table}%
This agrees with the results of the wavelet analysis within error. For the radio emission at 2.5 MHz, this analysis was done for the four main peaks in the time-series, shown in green in Figure 7 as well as seven peaks which include lower amplitude pulsations, shown in blue in Figure 7. This resulted in periods of $\sim 230$ s and $\sim 157$ s, respectively. Therefore, this result matches well with the wavelet analysis when only the four main peaks are accounted for. When the less significant peaks are included, we see that the period draws closer to that of the higher frequency radiation. The matching time-scales of the 171~\AA, 1600~\AA, SXR, and HXR light curves indicate that the mechanism producing the QPPs in these wavebands must have the same progenitor, that is also likely related to the radio emission observed. Our interpretation of these results and the relationship between the emission in each waveband is detailed in Section 4.
\\
\\
\\
\\
\\
\\
\\
\\
\subsection{Spatial analysis}
To investigate spatially the regions of the flare from which the QPPs originate, we conducted imaging analysis using RHESSI and SDO/AIA. No radio imaging instrument was available during the observation. Firstly, we used the PIXON algorithm to determine where the non-thermal HXRs originated from. The imaging was carried out over an energy band of 35-70 keV during the available time period when RHESSI was observing the event before entering spacecraft night: $\sim$13.35-13.42 UT. This includes the first two prominent pulsations in the sequence of 7. The light curve showing the available RHESSI data is shown in Figure 9 in magenta. It was found that there were three HXR sources on the map which are labelled within three kernels as K1, K2, and K3. The ribbons of the flare are clearly in visible in Figure 9a. K1 and K2 lie along the higher ribbon while K3 is located on the lower ribbon. A system of flare loops connects these ribbons. Figures 9a and 9e show these sources in red overlaid on top of the 1600~\AA\ and 171~\AA\ backgrounds, respectively. The event occurred close to the disk centre and had a loop footpoint separation of $\sim$50 Mm and a loop height of $\sim$25 Mm. This was estimated by measuring the separation of the HXR footpoints and assuming a semi-circular geometry of the loops.

These HXRs are produced through non-thermal bremsstrahlung through interaction of the flare accelerated electrons with the dense chromosphere which acts a `thick-target' \citep{brown}. The mechanism modulating the HXRs that produces the observed QPPs must be causing a sequence of episodic or `bursty' energy releases that intermittently accelerates electrons resulting in a modulation of the non-thermal bremsstrahlung emission. We discuss this further and its relevance to the QPPs in the other wavebands in Section 4. 
\begin{figure}[t!]
\centering
\includegraphics[width=8.5cm]{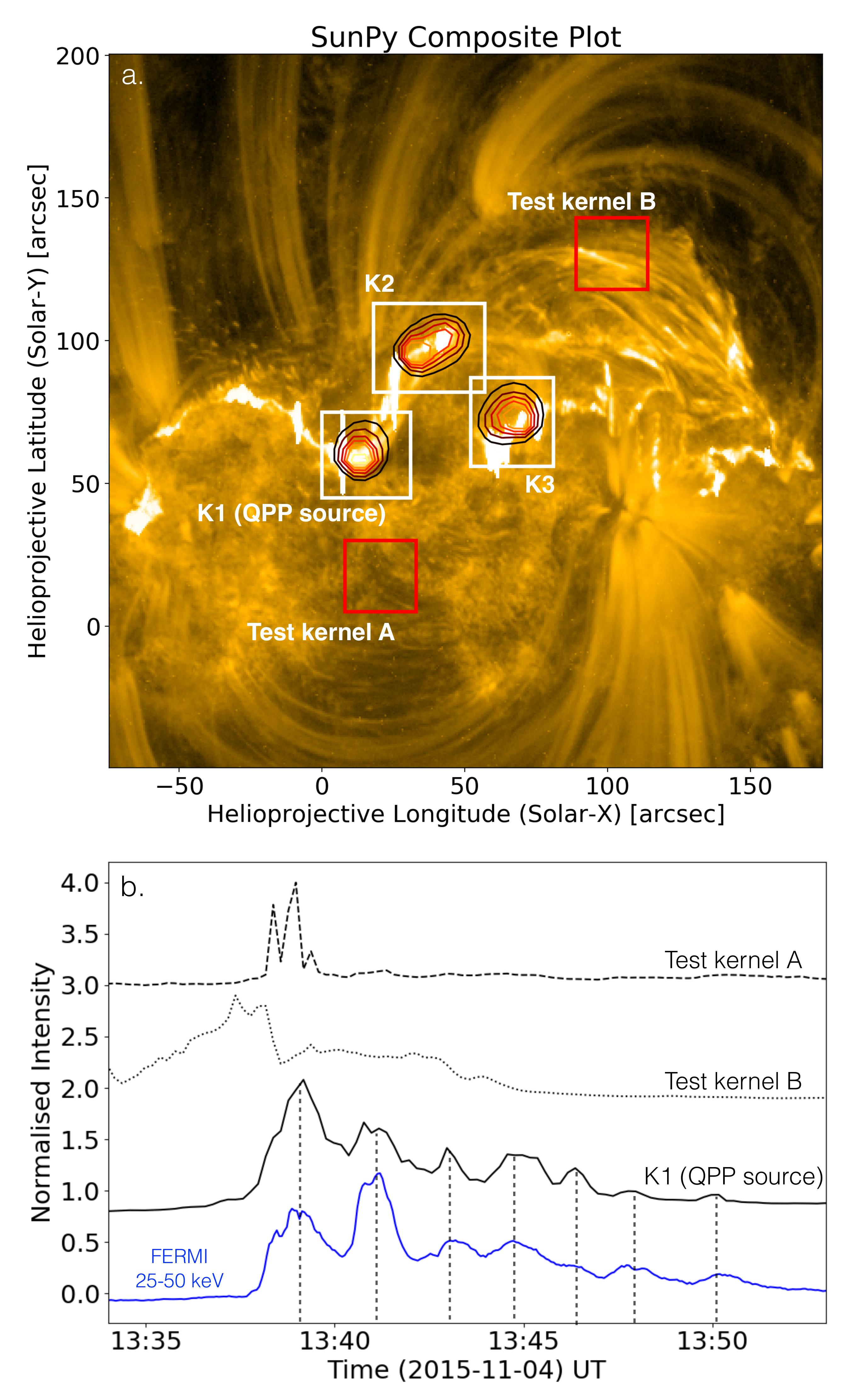}
\caption{(a): SDO/AIA 171~\AA\ passband image of the flare site on 2015 Nov 4. Shown are the QPP source (within K1), the additional two HXR sources (within K1 and K2), and two test kernels representing the arbitrary regions of the map. (b): The time-series obtained from K1 (QPP source) compared against those obtained from the test kernels. It is clear that the time-series constructed using the test kernels are uncorrelated to the HXR QPPs while the light curve obtained from K1 matches the HXR profile.}
\end{figure}

To determine the location within the flaring region producing the QPPs in the 171~\AA\ and 1600~\AA\ emission, we created time-series from the images taken from SDO/AIA. To localise the QPP source, we integrated the emission from each image over each region of the entire active region using various kernel sizes, generated time-series for each of these kernels for the duration of the flare, and compared the profiles of the time-series to that of the HXR emission. This allowed us to obtain the flux from within each test kernel at each time-step to compare to the HXR emission. It was found that the kernel that produced the most prominent QPPs, as well as having the same characteristic periodicity as the HXR QPPS, spatially coincided with the location and size of the HXR source at K1. Figure 9b and 9f illustrate this in that there is a strong correlation between the QPPs in the EUV emission extracted from K1 and the HXR emission. The EUV emission from K2 and K3 is significantly less correlated to the HXR emission as shown in Figures 9c, 9d, 9g, and 9f. 

The animation associated with Figure 9, available in the online version of this article, shows the evolution of the flare at each time step. It is clear that there is an asymmetry between the light curves obtained for the EUV emission at each HXR source location with K1 being the most correlated to the HXR QPPs. This analysis suggests that K1 is the region of the flare site in which the QPPs originate. Figure 10 shows an additional comparison of the EUV emission from K1 (QPP source) and the emission obtained from two test regions not associated with the HXR sources. Here, we can see again that integrating each time step over K1 produces QPPs highly correlated with the HXR emission while doing so for each test kernel does not. This trend continues no matter which region of the flare is used to construct the EUV time-series. 

K1 is associated with open magnetic field lines, identified in the potential field source surface extrapolation (PFSS) shown in Figure 11. PFSS models provide an approximation of the coronal magnetic field up to 2.5 $R_\odot$ based on the observed photospheric field (\citealp{Schrijver2003}). Here the PFSS is calculated using \texttt{pfsspy} \citep{Stansby2020}. This magnetic field geometry allows for a mechanism for the escape of the electrons responsible for producing the radio emission. In the following section, we discuss the interpretation of these data and what proposed models of QPP generation allow for these observations.
\begin{figure*}[t!]
\centering
\includegraphics[width=13cm]{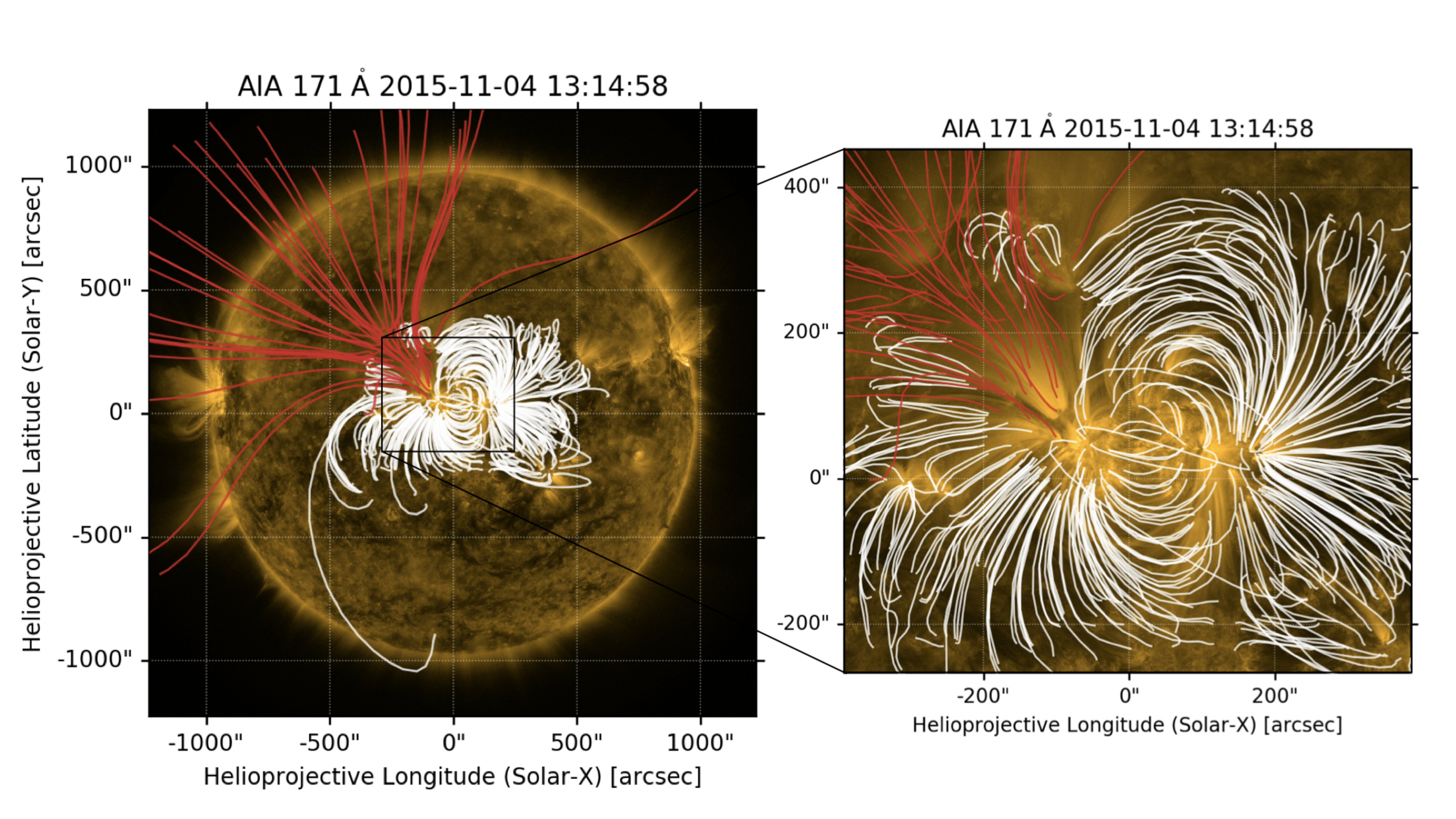}
\caption{Potential Field Source Surface (PFSS) extrapolation showing the geometry of the magnetic field lines of the flaring region overlaid on the AIA 171~\AA\ image. The open field lines are plotted in red and the closed lines in white. It is clear that the K1 has an open field line source and we propose that the interaction between the closed and open field lines at this footpoint result in `bursty' magnetic reconnection giving rise to the QPPs we observe. The open field lines allow the flare-accelerated electrons to escape that produce the Type \rom{3} radio emission.}
\end{figure*}
\section{Discussion and conclusions}\label{discussion}
Our results indicate that the EUV (171~\AA\ and 1600~\AA), SXR, and HXR QPPs contain approximately the same periodicity. We also observe QPPs in the low frequency radio domain in the form of a sequence of type \rom{3} radio bursts that occur during the time of the flare that have a longer periodicity. Our spatial analysis suggests that the EUV and HXR QPPs originate from the same region of the flare - the HXR footpoint at K1. This points towards a scenario in which intermittent particle acceleration is occurring due to a process inducing time-dependent magnetic reconnection. This particle acceleration occurs in a quasi-periodic fashion and results in bursty non-thermal bremsstrahlung that modulates the HXR emission occurring at the footpoints. The EUV emission would then be a consequence of this process as the ambient plasma is heated as the precipitating accelerated particles lose their energy.

The asymmetry of the EUV pulsations present at each HXR source, as shown in Figure 9, suggests that the electrons from the reconnection site must be preferentially accelerated between the closed loops and open field lines close to K1. Figure 11 shows these systems of open and closed field lines obtained via a PFSS extrapolation. It is likely that the radio emission observed is a consequence of the same intermittent particle acceleration that resulted in the EUV and HXR pulsations. However, the electrons accelerated along the open magnetic field lines from flare region result in the radio emission while it is the precipitating electrons accelerated towards the chromosphere which result in the HXR/SXR/EUV emission. Unfortunately no imaging observations at these radio frequencies are available during this event, and so we could not image the radio source to localise its origin. However, there are no nearby active regions at the time of the flare that could have coincidentally produced this radio emission.

To explain our observations we interpret the QPPs identified in this flare in terms of pulsed electron acceleration caused by time-dependent intermittent reconnection. In Figure 12 we show a cartoon scenario of the flare site to illustrate how the QPP sources are related to the magnetic field configuration. Following each burst of electron acceleration, those that escape upwards along the open magnetic field lines result in the type \rom{3} QPPs, and those that travel along closed lines precipitate in the chromosphere to cause the QPPs we observe in hard X-ray and EUV. But what causes the reconnection and particle acceleration itself to be quasi-periodic? As mentioned in the introduction, this could be due to either the process itself being time-dependent (self-oscillatory) or indeed due quasi-periodic triggering of magnetic reconnection due to external MHD waves. We can rule out the latter, as it is unlikely as there are no active regions nearby.

Plasmoid magnetic island reconnection or oscillatory reconnection are both good candidates. Given that the period of the QPPs analysed in this work match well with the simulations in \cite{McLaughlin2012Oscillatory} (105-121.5 s), this mechanism may be responsible. \cite{McLaughlin2012Oscillatory} outline how the interaction of magnetic flux emerging from the tachocline with an existing magnetic topology such as a flaring system can result in oscillatory reconnection and pulsed particle acceleration. It is possible that this flux emergence is localised to the region of the flare site we identified as the QPP source. This could then give rise to the QPPs we observe across the electromagnetic spectrum. However we are unable to rule out the possibility of plasmoid magnetic island reconnection or other self-oscillatory processes. There are a number of arguments that point towards a relation between the HXR/SXR/EUV QPPs and the radio QPPs we observe despite them having different periods according to our wavelet analysis. We outline below our argument that they are indeed a consequence of the same progenitor.

1. The emission mechanism involved that produces the radio (plasma emission) versus the mechanism producing the EUV, SXR, and HXR (non-thermal bremsstrahlung/heating) are very different in nature - i.e incoherent free-free emission versus coherent collective emission. In the plasma emission mechanism, accelerated electron beams travel to large heights (for the frequencies we observe) along open magnetic field lines, induce the growth of Langmuir waves, and then these Langmuir waves must interact to finally produce radio emission \citep{melrose2017}. 
Many factors, such as the electron energy (which can vary from pulse to pulse), velocity dispersion, Coulomb collisions, Langmuir wave growth and interaction, to name a few, play a role in generating the emission. 
It is a multi-stage process, and variability in any of these stages can change the characteristics of the radio pulses.
In contrast, the electrons producing the higher energy radiation, via non-thermal bremsstrahlung and subsequent heating of the surrounding plasma, must only travel from the acceleration site within the flaring region to the chromosphere. Bremsstrahlung then occurs quickly followed by instantaneous heating resulting in co-temporal pulsations in the EUV, SXR, and HXR wavebands \citep{xray}. Due to these factors, it is expected that not every HXR/SXR/EUV pulsation would have a corresponding radio burst, as we observe, despite being a consequence of the same intermittent particle acceleration.
\begin{figure}[t!]
\centering
\includegraphics[width=8.5cm]{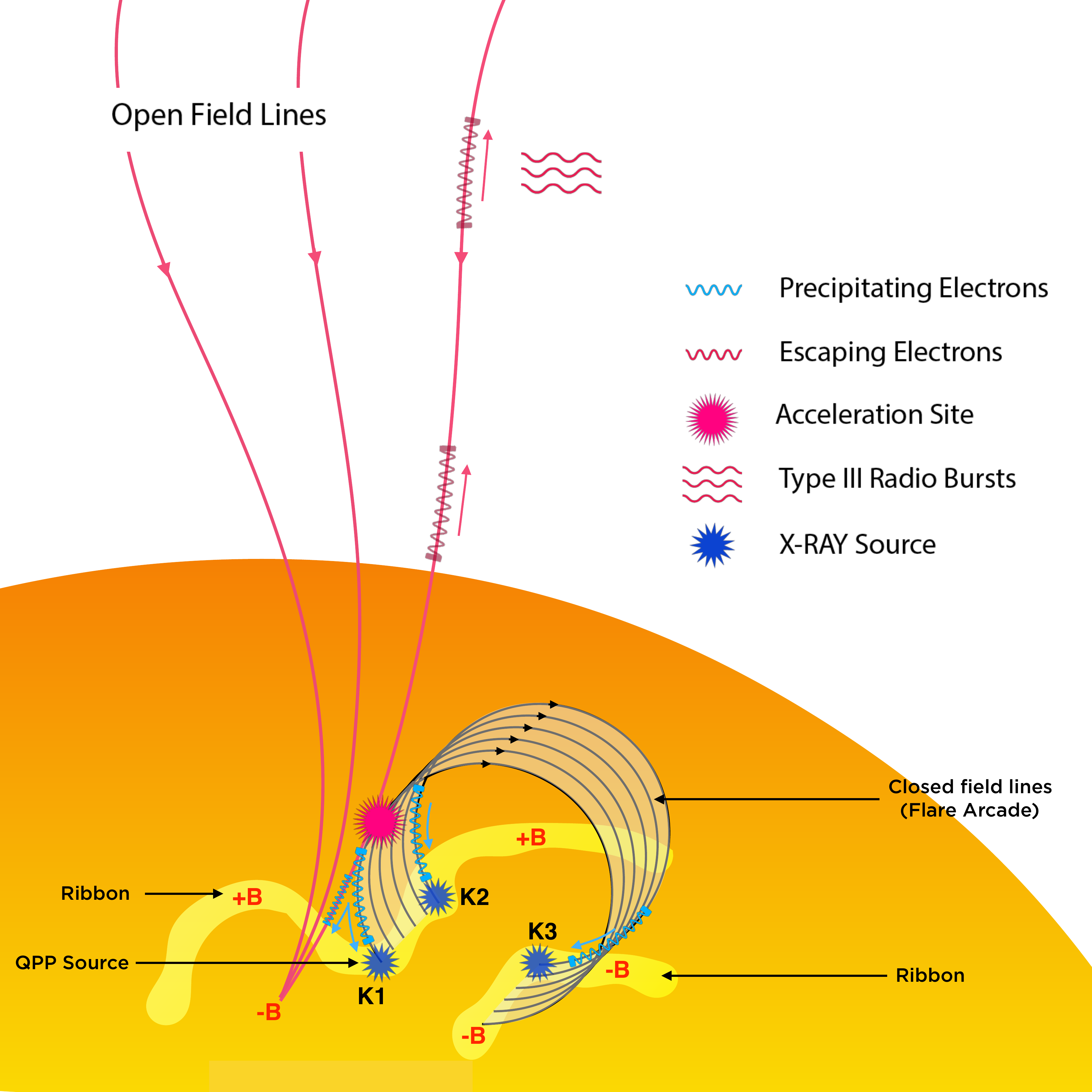}
\caption{Cartoon of the flaring region illustrating the likely mechanism through which we observe the episodic particle acceleration resulting in QPPs in EUV, radio, SXR, and HXR. The QPP source footpoint is related to the open and closed field lines allowing for the escape of the electrons resulting in the radio emission and the precipitation of the electrons giving rise to the higher frequency emission.}
\end{figure}

2. The region of the flare site we have identified as the QPP source is in close proximity to open and closed magnetic field lines. This magnetic field geometry allows for reconnection to occur between the open and closed field lines, providing a natural route for the escaping electrons to produce the radio emission and the precipitating electrons to produce the X-rays/EUV as shown in Figure 11.

3. The time delay between the HXR emission and the onset of the prominent type \rom{3} radio bursts is consistent with the estimated distance over which the radio electron beam sources must travel before they emit at 2.5 MHz, as shown in Figure 3. This source height of ${\sim}16$ $R_\odot$ was obtained by the \cite{newkirk1967} electron density model.

4. The wavelet analysis of the 2.5 MHz light curve only picks up the four main peaks in the time series. Smaller amplitude peaks fail to contribute significantly to the result. In Figure 8 we manually find the period by identifying the four most prominent peaks, which matches the result of the wavelet analysis. However when the smaller amplitude peaks are accounted for, amounting to a total of 7 pulses, the period of the radio emission comes within error of the period of the HXR/SXR/EUV. An additional difficulty in accurately calculating the period of the radio emission is that certain bursts are more intense at different frequencies as is clear in the dynamic spectra. However, from inspecting Figure 13 where the HXR emission is overplotted on the dynamic spectrum, there is quite a clear relation between the radio bursts and the HXR peaks when the entire frequency band is taken into account.
\begin{figure}[b!]
\centering
\includegraphics[width=8cm]{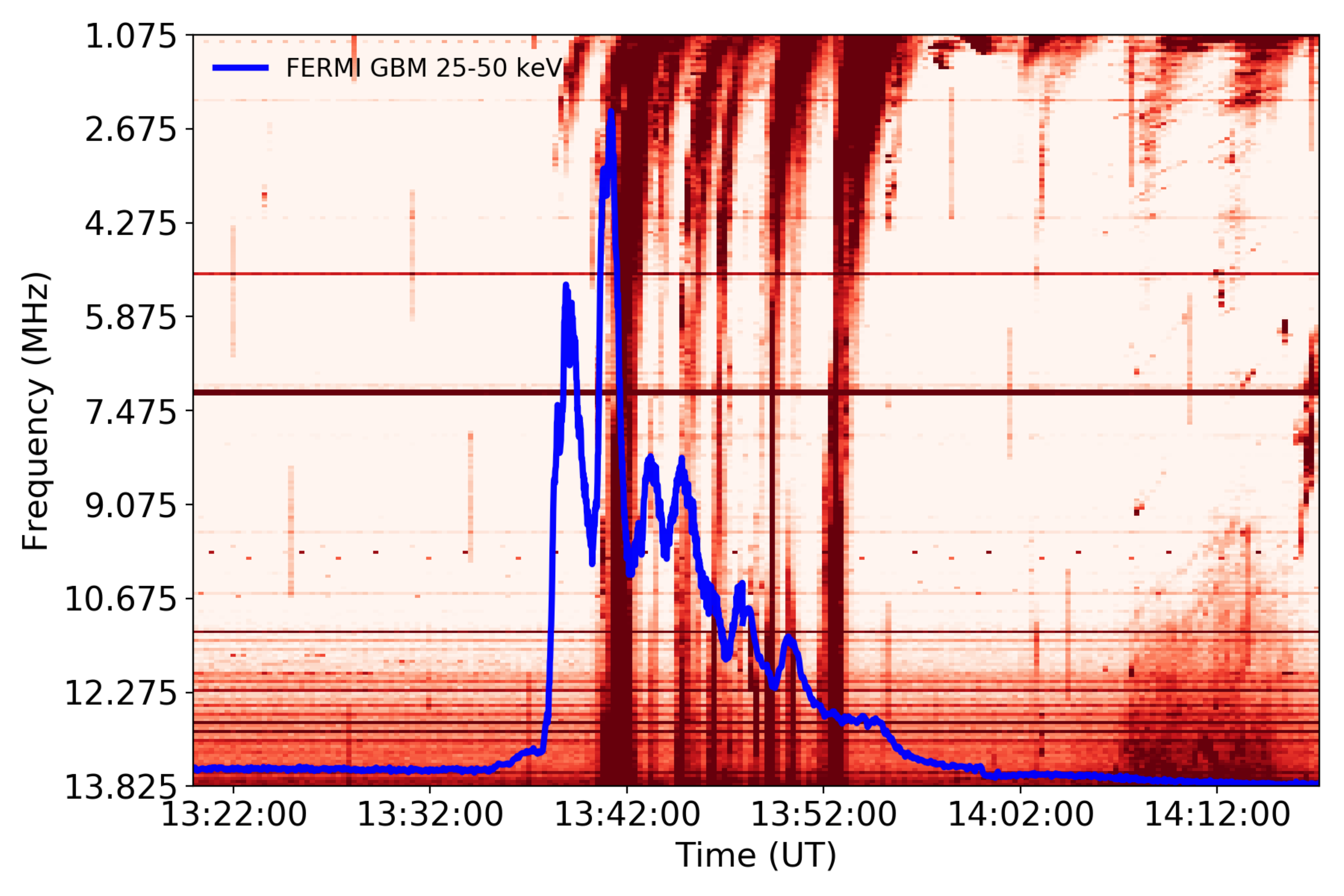}
\caption{The HXR emission from FERMI GBM (25-50 KeV) overplotted against the dynamic spectrum from WIND/WAVES showing the low frequency radio emission in the form of a sequence of type \rom{3} radio bursts.}
\end{figure}

\cite{cairns2020} point out that a type \rom{2} radio burst occurs at the time of this flare and suggest that the associated shock may be responsible for accelerating the electrons that result in the low frequency radio emission. However, considering the arguments above (points 1-4), we conclude that it is more likely that the type \rom{3} radio bursts are due to pulses of electron beams accelerating along the open magnetic lines close to the QPP source region. Additionally, the dynamic spectra of the radio emission from kHz to GHz shows traces of type \rom{3} bursts that extend to high frequencies, above the frequency of the type \rom{2} burst (see \cite{cairns2020} Figure 15). This suggests that they originate from a region closer to the flare site.

In summary, A multi-wavelength analysis of QPPs in an M-class flare has been conducted. Several instruments were used to allow for the analysis of the HXR, SXR, EUV, and radio emission detected during the event. The 171~\AA, 1600~\AA, SXR, and HXR light curves yielded similar periods of $122^{+26}_{-22}$ s, $131^{+36}_{-27}$ s, $123^{+11}_{-26}$ s, and $137^{+49}_{-56}$ s, respectively, indicating a common underlying mechanism, while the radio emission at 2.5 MHz contained a longer period of ${\sim}231$ s. X-ray and EUV imaging enabled us to localise the QPP source to a region of the flare site associated with open magnetic field lines. We found that the time delay between the X-ray/EUV emission and the radio emission is consistent with the estimated distance over which the electron beam sources must travel. We discuss the differences between the emission mechanisms responsible for the HXR/SXR/EUV emission versus the radio emission and determine that the QPPs in each waveband are linked to the same populations of accelerated electrons. We conclude that the QPPs in this event are due to some time-dependent self-oscillatory reconnection mechanism. Magnetic reconnection occurring in this bursty fashion injects populations of non-thermal electrons into the flare site giving rise to the sequence of pulses we observe in the SXR, HXR, and EUV as electrons collide with the chromosphere while the electrons accelerating away from the flare site along open magnetic field lines produce the type \rom{3} radio bursts. This work provides new evidence that oscillatory reconnection can naturally generate quasi-periodic periodic pulsations providing an explanation for their presence across the entire spatial range of flaring emission. This work also shines light onto the nature of energy release in flares and provides new insight into how QPPs may be localised to specific regions within flare sites. Future work that investigates the details and conditions required for the triggering of magnetic reconnection in this bursty fashion is needed. 
\acknowledgments
{This work has been supported by the European Space Agency PRODEX Programme (BPC) and the Government of Ireland Studentship from the Irish Research Council. L.A.H. is supported by an appointment to the NASA Postdoctoral Program at Goddard Space Flight Center, administered by USRA through a contract with NASA. We also thank the anonymous referee whose comments helped to improve this paper.}
\vspace{5mm}
\facilities{SDO (AIA), RHESSI, WIND (WAVES), FERMI (GBM),GOES/XRS}

\software{sunpy \citep{sunpy_community2020, Mumford2020}
pfsspy \citep{Stansby2020}
matplotlib \citep{Hunter:2007}}
\clearpage
\appendix
\section{Wavelet analysis without detrending technique}
\begin{figure*}[h]
\centering
\includegraphics[width=15cm]{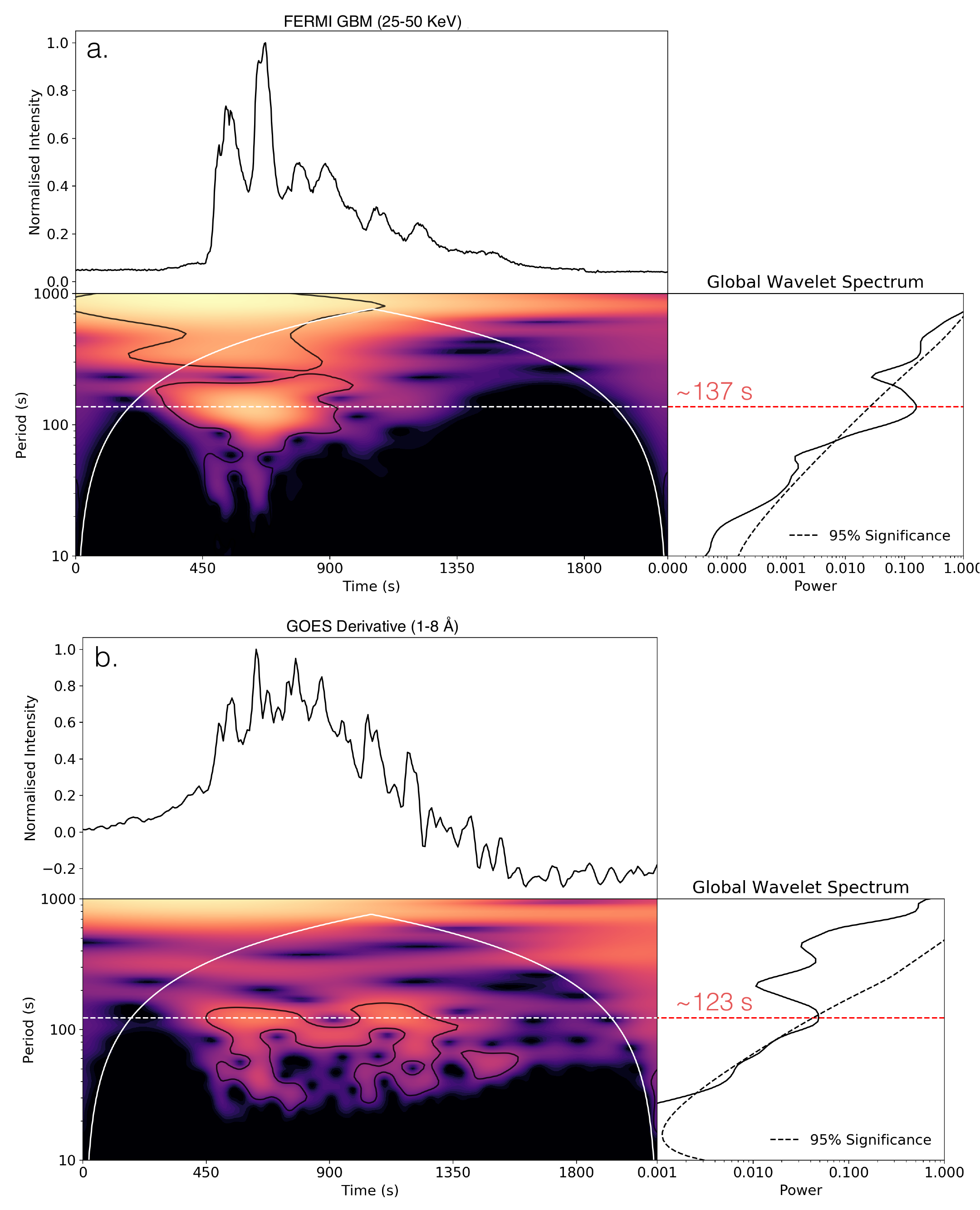}
\caption{Wavelet analysis of the (a): HXR and (b): SXR derivative emission from the flare without detrending. The periods were found to be $137^{+64}_{-61}$ s and $123^{+6}_{-34}$ s, respectively. The error is taken as the range over which each global power spectrum is above 95\% significance.}
\end{figure*}

\begin{figure*}[h]
\centering
\includegraphics[width=15cm]{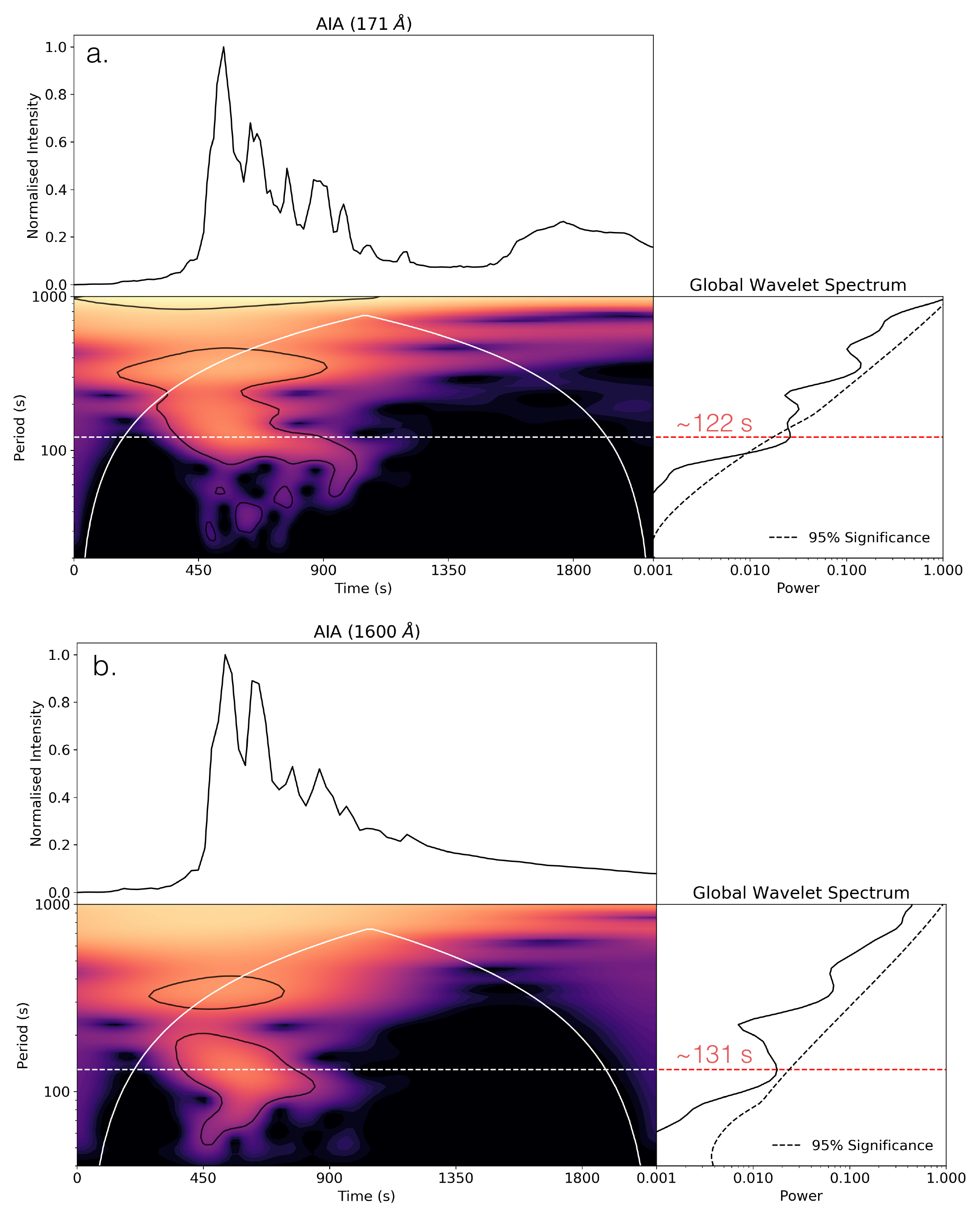}
\caption{(a): Wavelet analysis without detrending of the emission at 171~\AA. (b): Wavelet analysis without detrending of the emission at 1600~\AA. The periods were found to be $122^{+17}_{-26}$ s and ${\sim}131$ s respectively.}
\end{figure*}

%

\bibliography{sample63}{}
\bibliographystyle{aasjournal}



\end{document}